\documentclass[journal]{IEEEtran}

\usepackage{graphicx}
\usepackage{cite}
\usepackage[cmex10]{amsmath}
\usepackage{array}

\ifCLASSOPTIONcompsoc
 \usepackage[caption=false,font=normalsize,labelfont=sf,textfont=sf]{subfig}
\else
 \usepackage[caption=false,font=footnotesize]{subfig}
\fi

\usepackage{stfloats}

\ifCLASSOPTIONcaptionsoff
 \usepackage[nomarkers]{endfloat}
\let\MYoriglatexcaption\caption
\renewcommand{\caption}[2][\relax]{\MYoriglatexcaption[#2]{#2}}
\fi

\usepackage{url}

\begin{document}
% \title{Metal artifact reduction via artifacts correlation}
\title{Metal artifact reduction based on the combined prior image}

%\author{Michael~Shell,~\IEEEmembership{Member,~IEEE,}
%        John~Doe,~\IEEEmembership{Fellow,~OSA,}
%        and~Jane~Doe,~\IEEEmembership{Life~Fellow,~IEEE}% <-this % stops a space
%\thanks{M. Shell is with the Department
%of Electrical and Computer Engineering, Georgia Institute of Technology, Atlanta,
%\thanks{Manuscript received April 19, 2005; revised December 27, 2012.}}

\author{Yanbo~Zhang, Xuanqin~Mou% <-this % stops a space

\thanks{This paper was presented on the 12th  International  Meeting  on  Fully  Three-dimensional  Image  Reconstruction  in
Radiology and Nuclear Medicine in 2013. It was submitted on 3 February 2013; accepted on  21 March 2013; orally presented on 20 June 2013. The conference proceedings can be downloaded from the website http://www.fully3d.org/. This paper is cited as "Yanbo  Zhang  and  Xuanqin  Mou.  Metal  artifact  reduction  based  on  the  combined  prior  image. Proceedings  of  the  12th  International  Meeting  on  Fully  Three-dimensional  Image  Reconstruction  in Radiology and Nuclear Medicine, Lake Tahoe, California, 2013, pp. 412-415."}

\thanks{This work was partly supported by the NSFC through Grant No. 61172163 and the Research Fund for
the Doctoral Program of Higher Education of China  through Grant No. 20110201110011.}% <-this % stops a space

\thanks{Yanbo Zhang and Xuanqin Mou are with the Institute of Image
Processing and Pattern Recognition,  Xi'an Jiaotong University, Xi'an, Shaanxi 710049, China. (Email: yanbozhang007@163.com, xqmou@mail.xjtu.edu.cn).}

%(corresponding author: Xuanqin
%Mou, phone: 86-29-82663719; e-mail: xqmou@mail.xjtu.edu.cn)

%\thanks{M. Shell is with the Department
%of Electrical and Computer Engineering, Georgia Institute of Technology, Atlanta,

%\thanks{Manuscript received April 19, 2005; revised December 27, 2012.}

}

\maketitle
\begin{abstract}
Metallic implants introduce severe artifacts in CT images, which degrades the image quality. It is an effective method to reduce metal artifacts by replacing the metal affected projection with the forward projection of a prior image. How to find a good prior image is the key of this class methods, and numerous algorithms have been proposed to address this issue recently. In this work, by using image mutual correlation, pixels in the original reconstructed image or linear interpolation corrected image, which are less affected by artifacts, are selected to build a combined image. Thereafter, a better prior image is generated from the combined image by using tissue classification. The results of three patients' CT images show that the proposed method can reduce metal artifacts remarkably.

\end{abstract}


% \fi
\IEEEpeerreviewmaketitle

\section{Introduction}
% \IEEEPARstart{T}{his}

\IEEEPARstart{M}{etal} artifact reduction (MAR) is a major problem in x-ray computed tomography. Metallic implants can introduce bright and dark streaks and shadows in CT images, which degrades the image quality severely and become a major limiting factor in clinical diagnosis. During past three decades, various metal artifact reduction approaches have been proposed. However, there is still no robust solution to this issue and it remains a challenging problem.
% \cite{Abdoli2012review}

The projections passing through metals are distorted by various errors such as severe beam hardening and noise \cite{Philips2012}. As a result, many MAR methods treat metal affected projections as to be missing, and they are replaced by surrogate projections. Some methods complete the projection by using an interpolation scheme, e.g., linear interpolation \cite{1987LIMAR} (denoted as LI-MAR), which is simple and with low computation cost. However, interpolation based MAR methods may introduce secondary artifacts.

Recently, a class of MAR methods, which complete the missing projection dataset by using forward projection of a prior image, are widely investigated. The information of the prior image is exploited to complete the projection, as a result, the method can get an excellent result if the prior image contains few artifacts and is close to the ground truth image. Therefore, how to find a good prior image is crucial in these forward projection based methods. Generally, a prior image is generated from the original reconstructed image or pre-corrected image. Seemeen Karimi et al. \cite{Seemeen2012} obtained the prior image by segmenting regions of the original image. Metal artifacts regions were identified and then replaced with a constant soft tissue value. Bal and Spies \cite{Bal2006} employed the k-means cluster technique to segment the adaptively filtered image into five classes. Prell et al.\cite{Prell2009} segmented three dimensional interpolation corrected image into air, soft tissue and bone. Philips Healthcare recently developed a commercial orthopedic metal artifact reduction function (O-MAR)
 %designed for radiation therapy
which produced the prior image from the original image \cite{Philips2012}. Meyer et al. \cite{Meyer2010} produced prior image from different images depending on the strength of existing artifacts. The original image was chosen to generate prior image in the case of existing minor artifacts, otherwise, LI-MAR corrected image was selected instead. There are two main drawbacks for these methods. In some cases, there exist wrong tissue classification due to severe artifacts. Besides, pixel values in bone remain unchanged because they vary over a large range; as a result, the artifacts in bone remain. These factors result in generating poor prior image and finally lead to dissatisfactory correction performance.

All the above mentioned methods employ the information in only one image, while we try to make the best of the information in both the uncorrected original image and LI-MAR corrected image.
% to generate a good prior image.
The distributions and intensities of artifacts are different in the original image and LI-MAR corrected image, so the pixels containing fewer artifacts in the two images are selected to build a combined image, which is used to generate a good prior image.
%The original image and LI-MAR corrected image are contaminated by metal artifacts and introduced secondary artifacts, respectively. The two images are divided into small patches, and we can find the one which contains fewer artifacts in the corresponding two patches by using image mutual correlation. Then the central pixels of selected patches are combined to build a new image, which is used to generate prior image.
Thereafter, the forward projection of the prior image is used to complete the projection dataset and the corrected image is reconstructed using FBP.

\section{Method}
The main idea of our approach is to generate the prior image that is obtained from the combination of original image and LI-MAR corrected image. The proposed method is composed of three steps: Metal traces segmentation, prior image generation and projection completion followed by image reconstruction.
\subsection{Metal traces segmentation}
In the original reconstructed image, metals are segmented out based on thresholding \cite{Prell2010}. Then the forward projection of the obtained metal only image is performed to get the metal traces, which specifies the projections affected by metals. These affected projections are replaced in the final step.

\subsection{Prior image generation}

\subsubsection{Linear interpolation}
For a given original sinogram $\textbf{p}$, whose the $i^{th}$ view $j^{th}$ bin pixel is denoted as $p_{i,j}$. In each view of the sinogram, e.g. in the $i^{th}$ projection view, if the projections $p_{i,k}$ and $p_{i,k+\Delta+1}$ are unaffected by metal, and the $\Delta$ projection pixels between them, $\{p_{i,j}|j\in[k+1,k+\Delta]\}$, are in metal trace, then the projections $\{p_{i,j}|j\in[k+1,k+\Delta]\}$ are replaced by the linear interpolation of $p_{i,k}$ and $p_{k+\Delta+1}$, which is stated as follows.
\begin{equation}
 \label{eq:LI}
 p^{LI}_{i,j}=p_{i,k} +\frac{p_{i,k+\Delta+1}-p_{i,k}}{\Delta+1}(j-k).
\end{equation}
Sometimes the unaffected projections are smoothed before interpolation in order to decrease the impact of noise. The unaffected projections remain unchanged in $\textbf{p}^{LI}$. Then the LI-MAR corrected image is reconstructed using FBP.

\subsubsection{Artifacts splitting}
\begin{figure}[tb]
   \centering
   \includegraphics[width = 0.4\textwidth]{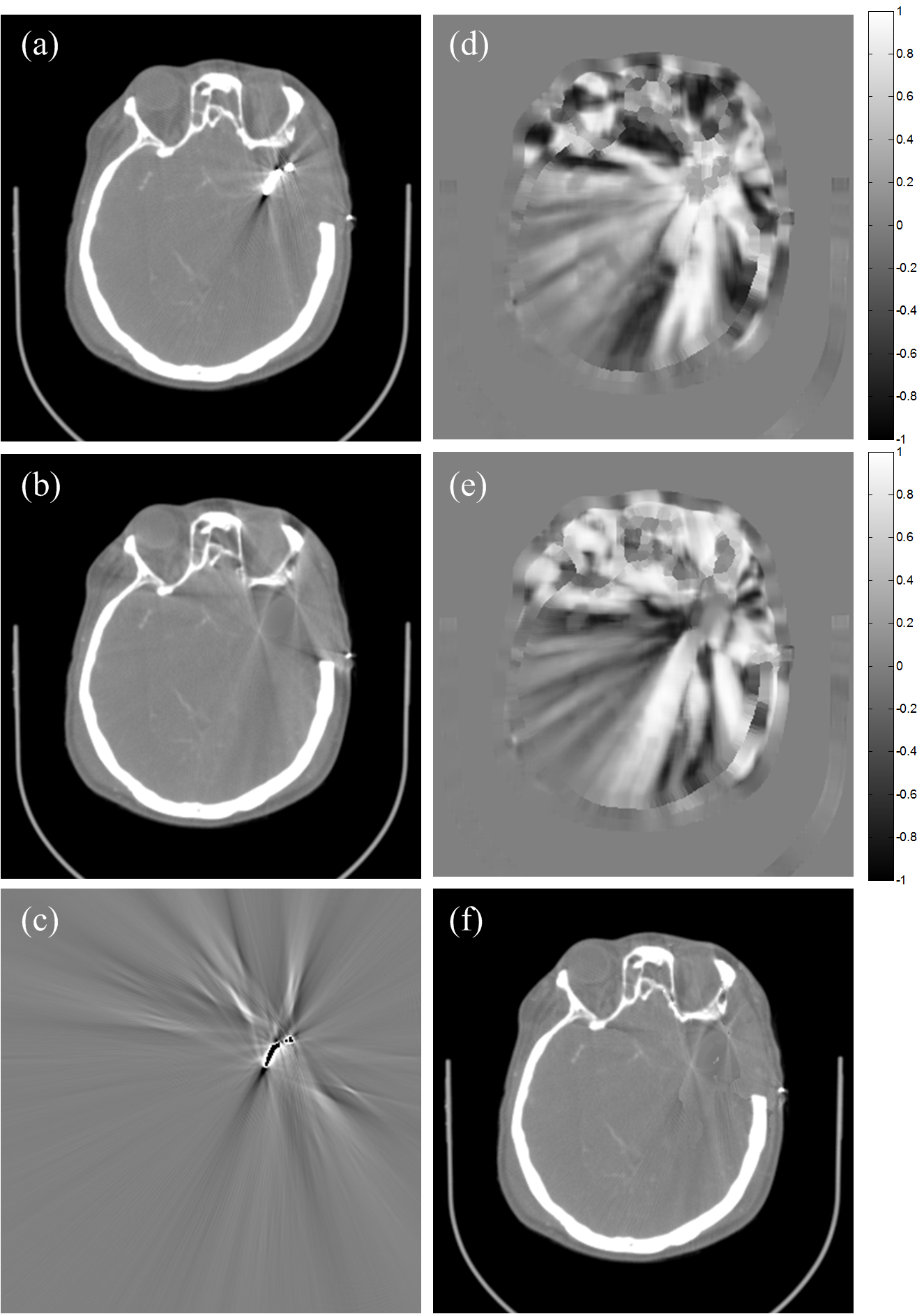}
   \caption{Generation of the combined images of patient 1. (a) is the original image, (b) is LI-MAR corrected image, (c) is the difference between (a) and (b), where the metal pixels are excluded. (d) and (e) are the correlation maps of (a) and (b), respectively. (f) is the combined image. The display windows are (WL=0HU, WW=1500HU) for (a--c) and (f), and [-1, 1] for (d) and (e).}
   \label{fig:combined}
  \end{figure}

The original image $\textbf{f}^O$ can be regarded as the ground truth image plus a metal artifacts image, and the LI-MAR corrected image $\textbf{f}^{LI}$ can be treated as the ground truth image plus a LI-MAR artifacts image consisting of uncorrected metal artifacts and secondary artifacts. Therefore, the difference of these two reconstructed images, $\textbf{f}^O-\textbf{f}^{LI}$, represents the superposition of metal artifacts and the negative LI-MAR artifacts, called as artifacts superposition image and denoted as $\textbf{f}^{A}$. Figure~\ref{fig:combined}(c) gives the artifacts superposition image of a patient (patient 1) containing a metallic clip. It can be seen that all metal artifacts and LI-MAR artifacts are presented and there is no information of patient' tissues.
%Figures 1(a) and 1(b) show that the distributions of metal artifacts and LI-MAR artifacts are different, so it is possible to automatically determine a given tissue pixel is
\subsubsection{Correlation maps}
Image mutual correlation \cite{Wang2004IQA} has been widely used to assess the degree of similarity of two images. The mutual correlation of two vectors $\textbf{x}$ and $\textbf{y}$ is calculated as follows:
\begin{equation}
 \label{eq:correlation}
 C(\textbf{x},\textbf{y})=\frac{2{\langle \textbf{x},\textbf{y}\rangle}+\epsilon}{{\|\textbf{x}\|}^2+{\|\textbf{y}\|}^2+\epsilon}.
\end{equation}
where symbol $\langle \cdot, \cdot \rangle$ represents the inner product, and $\epsilon$ is a small positive constant to make sure that the denominator is not zero. The value range of $C(\cdot, \cdot )$ is $(-1,1]$.

Generally, the distributions and intensities of metal artifacts and LI-MAR artifacts are different (e.g., Figs. 1(a) and 1(b)), so it is possible to build a combined image with fewer artifacts from the two images $\textbf{f}^O$ and $\textbf{f}^{LI}$. Specifically, for an arbitrary pixel position $(i,j)$, the pixel value of the combined image is selected as $f^O_{i,j}$ or $f^{LI}_{i,j}$, depending on that whose CT number is less affected by artifacts. For this purposes, two reconstructed images $\textbf{f}^O$ and $\textbf{f}^{LI}$ as well as the artifacts superposition image $\textbf{f}^{A}$ are divided into blocks (called image patches), and denoted as $\textbf{b}^{O}_{i,j}$, $\textbf{b}^{LI}_{i,j}$ and $\textbf{b}^{A}_{i,j}$, respectively, where the subscript $(i,j)$ is position index of the central pixel of the image patch. Each patch has $9\times9$ pixels. As known to us, CT number of soft tissue is around 0 HU; thus, in cases of heavy artifacts, the CT number is dominated by artifacts in soft tissue region in the reconstructed images. Therefore, in soft tissue region, correlation value between two corresponding patches in the reconstructed image and in artifacts superposition image describes the correlation degree of contained artifacts in this patch in the reconstructed image, which is calculated by Eq.~\ref{eq:correlationmap}.\footnote{The artifacts superposition image is defined by $\textbf{f}^A =\textbf{f}^O-\textbf{f}^{LI}$, so the negative sign is used in Eq.~\ref{eq:correlationmap} to guarantee the positive correlation of LI-MAR induced artifacts between two corresponding patches in $\textbf{f}^{LI}$ and $\textbf{f}^O-\textbf{f}^{LI}$.}
\begin{equation}
 \label{eq:correlationmap}
 \left\{ \begin{gathered}
C^{O}_{i,j} = C(\textbf{b}^{O}_{i,j},\textbf{b}^{A}_{i,j}), \hfill \\
C^{LI}_{i,j} = -C(\textbf{b}^{LI}_{i,j},\textbf{b}^{A}_{i,j}). \hfill \\
\end{gathered}  \right.
%\end{aligned}
\end{equation}
These correlation values obtained form Eq.~\ref{eq:correlationmap} compose correlation maps $\textbf{C}^O$ and $\textbf{C}^{LI}$. Since the artifacts is continuous along the directions of streaks, correlation values in bone can be estimated by using neighbourhood interpolation. Then two correlation maps $\textbf{C}^O$ and $\textbf{C}^{LI}$ are obtained. Figures~\ref{fig:combined}(d) and ~\ref{fig:combined}(e) are the two correlation maps of patient 1.

\subsubsection{Generation of combined prior image}
If the corresponding correlation value of $f^O_{i,j}$ is smaller than that of $f^{LI}_{i,j}$, then $f^O_{i,j}$ is likely to contain fewer artifacts and is selected to build the combined image; otherwise, $f^{LI}_{i,j}$ is chosen. Thus the combined image $\textbf{f}^{C}$ is obtained according to Eq.~\ref{eq:combimg}. Figure~\ref{fig:combined}(f) shows the combined image of patient 1, which contains lighter artifacts than Figs.~\ref{fig:combined}(a) and~\ref{fig:combined}(b).
\begin{equation}
 \label{eq:combimg}
 f_{i,j}^C  = \left\{ \begin{gathered}
  f_{i,j}^O ,\quad \mathrm{if} \; C_{i,j}^O  < C_{i,j}^{LI} \hfill \\
  f_{i,j}^{LI} ,\quad others. \hfill \\
\end{gathered}  \right.
\end{equation}

The prior image is obtained via tissue classification of the combined image \cite{Prell2009}. The pixels with CT numbers larger than 200 HU are regarded as bone, which are unchanged; the pixels whose CT numbers are smaller than -600 HU, are assumed as air and set to -1000 HU; while the pixels with CT numbers between -600 HU and 200 HU are treated as soft tissue, which are uniformly set to 0 HU. Then the combined prior image is obtained.

\subsection{Projection completion and image reconstruction}
%In this step, if
If the metal trace pixels are directly replaced with the corresponding projections obtained by the forward projecting of the combined prior image, it may lead to discontinuity at the boundary of the metal traces, which introduces new streak artifacts. So it is necessary to prevent generating the discontinuity in replacement. Similar to our previous work \cite{Yanbo2013}, we apply linear interpolation again to generate a continuous transition $\textbf{p}^{T}$ between the forward projection $\textbf{p}^{C}$ and the original sinogram $\textbf{p}$.
%\begin{equation}
% \label{eq:LItrans}
% p^{trans}_{i,j}= (p_{i,k}-p^{cp}_{i,k})\\
%  + \frac{(p_{i,k+\Delta+1}-p^{cp}_{i,k+\Delta+1})-(p_{i,k}-p^{cp}_{i,k})}{\Delta+1}(j-k).
%\end{equation}
%\begin{equation}
%\label{eq:LItrans}
%\begin{aligned}
%p^{T}_{i,j}= (p_{i,k}-p^{C}_{i,k}) \qquad \qquad \qquad \qquad  \qquad \qquad \quad  \; \\
%  + \frac{(p_{i,k+\Delta+1}-p^{C}_{i,k+\Delta+1})-(p_{i,k}-p^{C}_{i,k})}{\Delta+1}(j-k).
%\end{aligned}
%\end{equation}
%Where the subscripts have the same meaning as in Eq.~\ref{eq:LI}.
The metal affected projections are replaced with the sum of $\textbf{p}^{C}$ and $\textbf{p}^{T}$. In this way, the affected projections can be completed seamlessly. Thereafter, the corrected image is reconstructed using FBP, and the metal obtained in the first step is inserted into the corrected image.
\begin{figure}[tb]
   \centering
   \includegraphics[width = 0.4\textwidth]{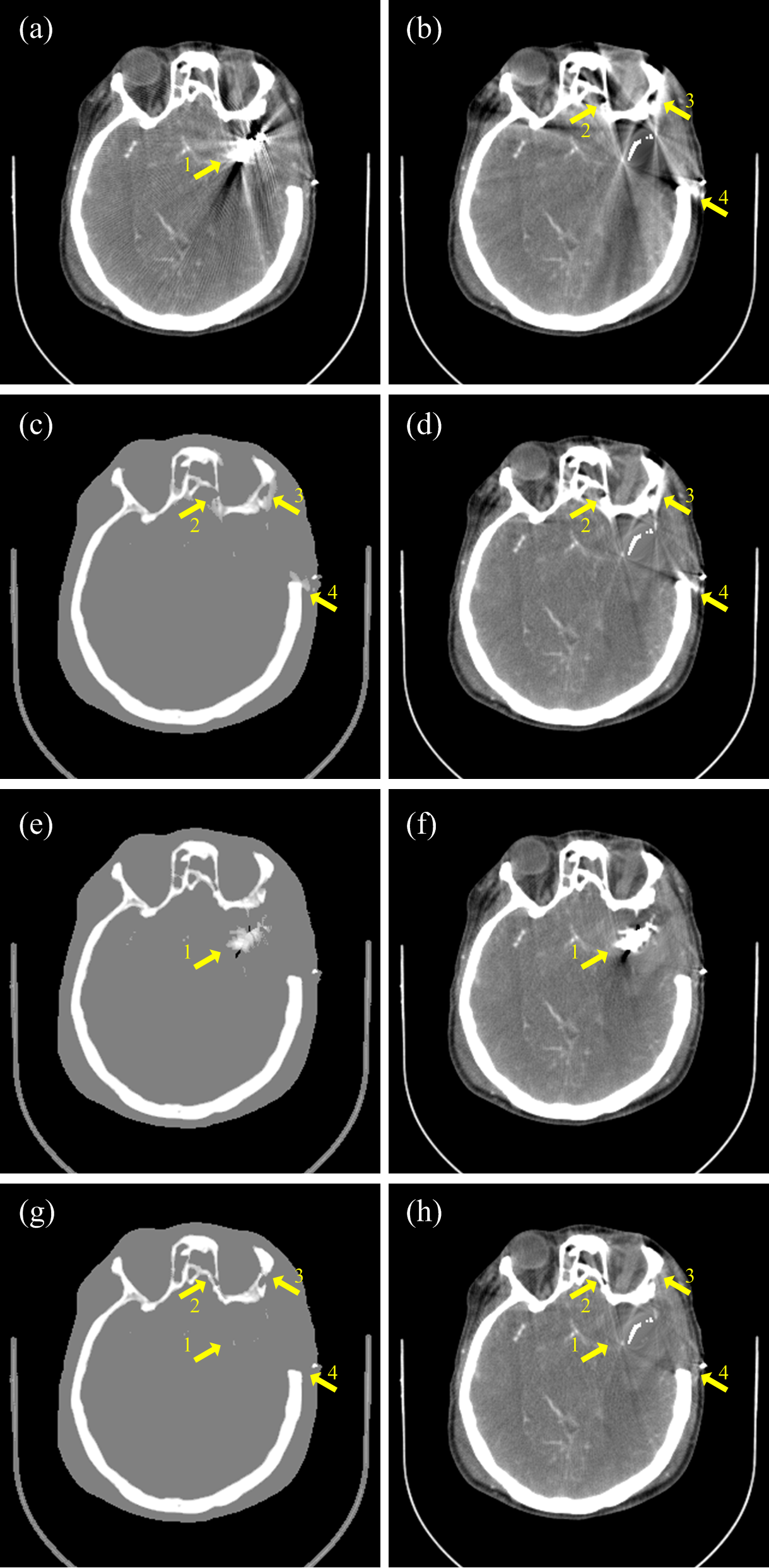}
   \caption{Reconstructed images and prior images of patient 1 with a surgical clip. (a) is the original image, the corrected images are obtained by using (b) LI-MAR, (d) FP-MAR1, (f) FP-MAR2 and (h) the proposed method, respectively. (c), (e) and (g) are the prior images of (d), (f) and (h), respectively. The display windows are (WL=50HU,WW=400HU) for reconstructed images and (WL=0HU,WW=1500HU) for prior images.}
   \label{fig:cliprecon}
  \end{figure}
\section{Experimental Results}
\subsection{Experimental setup}
%The acquired multislice dataset were converted into a stack of fanbeam sinograms by using rebinning method \cite{Noo1999}, and each sinogram associated with one horizontal z-slice.
In this study, we compare the performance of the proposed method with competing methods on the scanned datasets of three patients. A patient with a surgical clip (patient 1) was scanned on a Siemens SOMATOM Sensation 16 scanner CT using helical scanning geometry. The measurement of patient 1 was acquired with 1160 projection views over a rotation and 672 detector bins in a row. A patient with a dental filling (patient 2) and a patient with a hip prosthesis (patient 3) were scanned on a kV on-board imaging (OBI) system integrated in a $\mathrm{TrueBeam^{TM}}$ medical linear accelerator (Varian Medical System, Palo Alto, CA). The projection datasets were acquired with 364 projection views over 200$^\circ$ in full-fan mode for patient 2, and 656 projection views over 360$^\circ$ in half-fan mode for patient 3, respectively, and their effective detector bins were 512. The matrix of reconstructed image is $512\times512$, corresponding pixel sizes are $0.776mm\times0.776mm$ for patient 1 and patient 2, and $1mm\times1mm$ for patient 3.

%\begin{figure}[!t]
%\centering
%\includegraphics[width=2.5in]{pelvis_20130122.png}
% where an .eps filename suffix will be assumed under latex,
% and a .pdf suffix will be assumed for pdflatex; or what has been declared
% via \DeclareGraphicsExtensions.
%\caption{Simulation Results.}
%\label{fig_sim}
%\end{figure}

\begin{figure*}[tb]
   \centering
   \includegraphics[width = 1\textwidth]{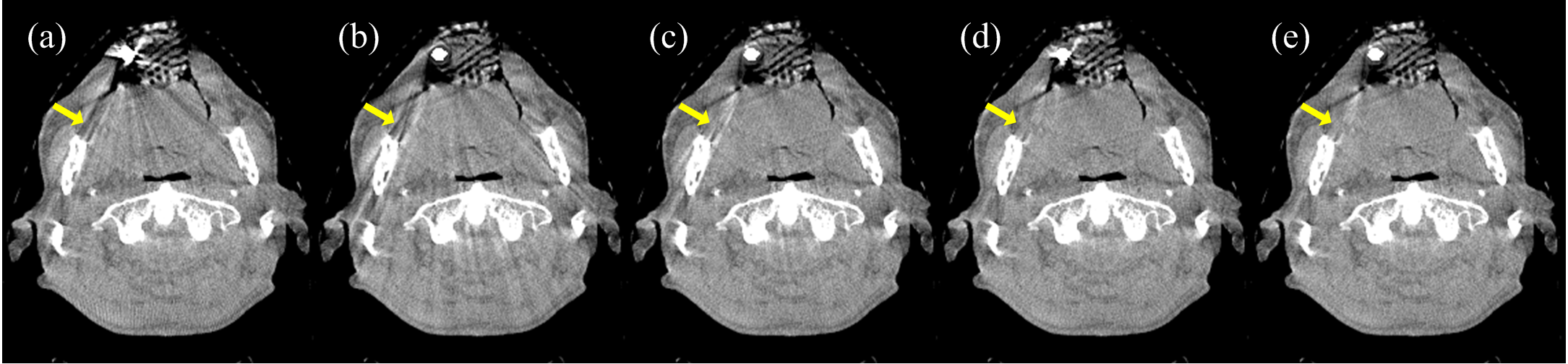}
   \caption{Reconstructed images of patient 2 with a dental filling. (a) is the original image, the corrected images are obtained by using (b)LI-MAR, (c) FP-MAR1, (d) FP-MAR2 and (e) the proposed method, respectively. The display window is (WL=0HU,WW=750HU).}
   \label{fig:dental}
  \end{figure*}
\begin{figure*}[tb]
   \centering
   \includegraphics[width = 1\textwidth]{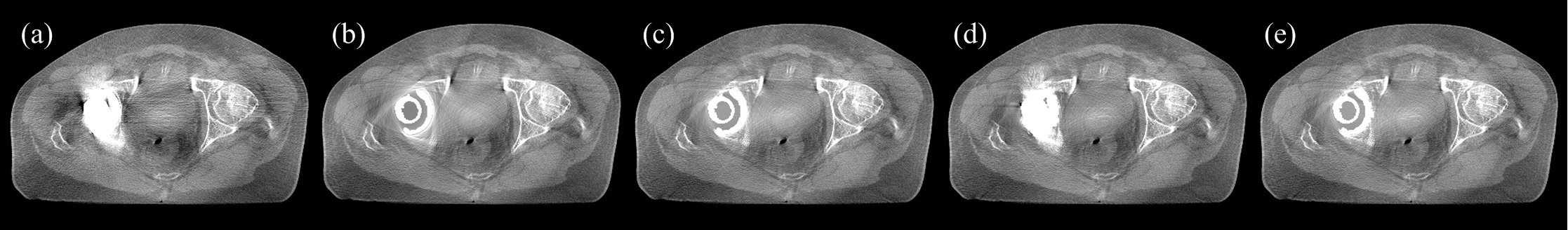}
   \caption{Reconstructed images of patient 3 with a hip prosthesis. (a) is the original image, the corrected images are obtained by using (b)LI-MAR, (c) FP-MAR1, (d) FP-MAR2 and (e) the proposed method, respectively. The display window is (WL=0HU,WW=750HU).}
   \label{fig:pelvis}
  \end{figure*}
\subsection{Results}
In this paper, the forward projection based MAR methods which use the prior images generated from the original image and LI-MAR corrected image are denoted as FP-MAR1 and FP-MAR2, respectively. FP-MAR1 is the same to Prell et al. proposed method~\cite{Prell2009} except that the conventional LI-MAR is used instead of 3D LI-MAR. LI-MAR, FP-MAR1 and FP-MAR2 are implemented to compare with the proposed method.

Figure~\ref{fig:cliprecon} shows the reconstructed images and the prior images of patient 1. The original image contains heavy metal artifacts in the vicinity of metallic clip as indicated by arrow 1 in Fig.~\ref{fig:cliprecon}(a). Thus the prior image obtained from the original image is not good enough due to wrong tissue classification as shown in Fig.~\ref{fig:cliprecon}(e), which results in severe artifacts in the corresponding reconstructed image (Fig.~\ref{fig:cliprecon}(f)). Figure~\ref{fig:cliprecon}(b) is the LI-MAR corrected image, which contains remarkable new artifacts as indicated by arrows 3 and 4, and the structure of bone is distorted highlighted by arrow 2. As a result, these artifacts and wrong structures are remained in the prior image (Fig.~\ref{fig:cliprecon}(c)) obtained from LI-MAR corrected image, leading to the similar artifacts and wrong structures as indicated by arrows in Fig.~\ref{fig:cliprecon}(d). By contrast, the combined prior image is obtained from Fig.~\ref{fig:combined}(f). Therefore, there is no previously mentioned wrong tissue classification as indicated by arrows in Fig.~\ref{fig:cliprecon}(g), and the corrected image has no obvious artifacts (see Fig.~\ref{fig:cliprecon}(h)).

Figure~\ref{fig:dental} shows the reconstructed images of patient 2. There are obvious streak artifacts in both the original image and LI-MAR corrected image. FP-MAR1 can remove most of streaks except the one pointed out by the arrow in Fig.~\ref{fig:dental}(c). In comparison, FP-MAR2 can suppress the streaks indicated by arrow better. Nevertheless, artifacts around the metal is remarkable in FP-MAR2 corrected image. The proposed method can reduce both streaks and artifacts around the metal greatly.

Figure~\ref{fig:pelvis} shows the reconstructed images of patient 3. There are bright and dark shadows in the original image, which can be reduced remarkably by LI-MAR. With the prior image obtained from LI-MAR, FP-MAR1 can further alleviate artifacts. On the contrary, FP-MAR2 can hardly reduce bright artifacts around the hip prosthesis. The image corrected by the proposed method is similar to that by FP-MAR1, because the two prior images obtained from LI-MAR image and combined image are both good enough.

\section{Discussions and Conclusions}

%For simplicity, LI-MAR method is adopted in this work. While of course, other MAR method can be used instead.
%
%The proposed method sufficiently exploit information of the better patches in original and LI-MAR corrected images. For some bone patches, both original and LI-MAR corrected ones are both contained heavy artifacts, the proposed approach is incapable to find a better one. This is based on the hyper........... In some cases, such as fixation screws inserted in a spine, LI-MAR blur the structure of spine around metals. As a consequence, the artifacts image, which is obtained from the difference of original image and LI-MAR corrected image, may contain the tissue structure information in the vicinity of metals. This may result in inaccuracy of mutual correlation between patches.

The distributions of artifacts in the original image and in LI-MAR corrected image are different, so the proposed method sufficiently exploits information of the pixels with fewer artifacts to compose a new image, which is used to generate a good prior image. As illustrated in the results of patient 1 and patient 2, images corrected by the proposed method are superior to that corrected by FP-MAR1 and FP-MAR2, because the combined prior image can avoid wrong tissue classification that appeared in the prior images of FP-MAR1 and FP-MAR2. For patient 3, since the prior image of FP-MAR1 has no wrong tissue classification and is superior to that of FP-MAR2, the prior image and corrected image of the proposed method are almost the same to that of FP-MAR1. Besides, for simplicity, LI-MAR method is adopted to generate the combined image in this work, while other MAR methods can also be used instead which will be our future work.

 %In some cases, such as fixation screws inserted in a spine, LI-MAR blur the structure of spine around metals. As a consequence, the artifacts image, which is obtained from the difference of original image and LI-MAR corrected image, may contain the tissue structure information in the vicinity of metals. This may result in inaccuracy of correlation maps.

In conclusion, we introduce a new method to generate better prior image for the forward projection based metal artifact reduction method. By using image mutual correlation, pixels in the original image or linear interpolation corrected image, which are less affected by artifacts, are selected to build the combined image. Based on this image, a more accurate prior image can be obtained. The results demonstrate that the proposed method can achieve better artifacts removal performance than the competing methods. In the future, the developed method will be evaluated by clinicians to validate the clinical usefulness.

% or


% -------------------------------
%\appendices
%% you can choose not to have a title for an appendix
%% if you want by leaving the argument blank
%\section{}

\section*{Acknowledgment}

The authors would like to thank Dr. Hao Yan, Dr. Xun Jia and Dr. Steve B. Jiang for providing the datasets with the dental filling and the hip prosthesis, and thank Dr. Hengyong Yu for providing the dataset with the clip.

% for his help on mutual correlation.
%
%
%% Can use something like this to put references on a page
%% by themselves when using endfloat and the captionsoff option.
%\ifCLASSOPTIONcaptionsoff
%  \newpage
%\fi
%



%\bibliographystyle{IEEEtran}
%\begin{thebibliography}{1}

\bibliographystyle{IEEEtran}

\bibliography{Fully3D_MAR}


%\end{thebibliography}


%\end{IEEEbiography}

%\newpage



%\vfill


\end{document}